\documentclass[twocolumn]{aastex62}
 
\usepackage{natbib}
\usepackage{graphicx}
\submitjournal{ApJL}

\shorttitle{Identifying DCBHs}
\shortauthors{Visbal and Haiman}

\begin{document}

\title{Identifying Direct Collapse Black Hole Seeds through their Small Host Galaxies}

\email{evisbal@flatironinstitute.org}

\author{Eli Visbal}
\affil{Center for Computational Astrophysics, Flatiron Institute, 162 5th Ave, New York, NY, 10010, USA}

\author{Zolt\'{a}n Haiman}
\affil{Columbia University, Department of Astronomy, 550 West 120th Street, New York, NY, 10027, USA}

\begin{abstract} 
Observations of high-redshift quasars indicate that super massive black holes (SMBHs) with masses greater than ${\sim}10^9~M_\odot$ were assembled within the first billion years after the Big Bang. It is unclear how such massive black holes formed so early. One possible explanation is that these SMBHs were seeded by ``heavy'' direct collapse black holes (DCBHs) with masses of $M_{\rm BH}\approx10^5~M_\odot$, but observations have not yet confirmed or refuted this scenario. In this Letter, we utilize a cosmological N-body simulation to demonstrate that before they grow roughly an order of magnitude in mass, DCBHs will have black hole mass to halo mass ratios much higher than expected for black hole remnants of Population III (Pop III) stars which have grown to the same mass (${\sim}10^6~M_\odot$). We also show that when $T_{\rm vir}\approx 10^4~{\rm K}$ halos (the potential sites of DCBH formation) merge with much larger nearby halos ($M_{\rm h} > 10^{10}~M_\odot$), they almost always orbit their larger host halos with a separation of a few kpc, which is sufficient to be spatially resolved with future X-ray and infrared telescopes. Thus, we propose that a future X-ray mission such as \emph{Lynx} combined with infrared observations will be able to distinguish high-redshift DCBHs from smaller black hole seeds due to the unusually high black hole mass to stellar mass ratios of
the faintest observed quasars, with inferred BH masses below ${\sim} 10^6 M_\odot$.
\end{abstract}


\section{Introduction}
Observations of high-redshift quasars imply that SMBHs with masses larger than ${\sim}10^9~M_\odot$ were assembled within the first billion years after the Big Bang \citep[e.g.][]{2011Natur.474..616M, 2015Natur.518..512W}. Understanding how these black holes grew to be so massive in such a short period of time is currently an unsolved problem \citep[for recent reviews see][]{2010A&ARv..18..279V, 2012RPPh...75l4901V, 2013ASSL..396..293H}. The simplest explanation would be that the first SMBHs were seeded by black hole remnants of Population III (Pop III) stars. However, a ${\sim} 100~M_\odot$ Pop III seed would need to grow at the Eddington limit for essentially the entire age of the Universe to reach the observed quasar masses, which seems unlikely due to feedback from nearby stars and the growing black hole seed itself \citep[e.g.][]{2009ApJ...701L.133A, 2009ApJ...696L.146M, 2018arXiv180406477S}.

One promising alternative is ``direct collapse'' black hole (DCBH) formation. In the DCBH scenario, ${\sim}10^5 ~M_\odot$ of gas quickly collapses into a black hole, possibly with a brief intermediate stage as a supermassive star \citep{1995ApJ...443...11E, 2002ApJ...569..558O, 2003ApJ...596...34B, 2006MNRAS.371.1813L, 2008MNRAS.391.1961D, 2014MNRAS.442.2036D, 2014MNRAS.445.1056V}. Larger seed masses have been shown to grow rapidly  \citep[e.g.][]{2008ApJ...676...33D, 2017ApJ...850L..42P}, more easily allowing the formation of the first SMBHs. In most models, DCBHs are predicted to form in $T_{\rm vir}\approx 10^4~{\rm K}$ ``atomic cooling'' dark matter halos (ACHs). DCBH formation likely requires the suppression of molecular hydrogen cooling to prevent fragmentation and star formation, which is usually is thought to occur as a result of strong H$_2$ dissociating radiation applied to a pristine ACH \citep{2001ApJ...546..635O, 2003ApJ...596...34B, 2010MNRAS.402.1249S}. However, additional channels have been proposed including high-velocity collisions \citep{2015MNRAS.453.1692I} and baryon-dark matter ``streaming velocities'' \citep{2010PhRvD..82h3520T, 2014MNRAS.439.1092T, 2017Sci...357.1375H, 2018MNRAS.479.4017I}.

Observationally determining the dominant SMBH seeding mechanism is very challenging. Currently detected high-redshift SMBHs are orders of magnitude more massive than ``heavy'' DCBH seeds and such a large amount of growth is likely to remove any memory of the initial seed mass. Thus, distinguishing DCBH seeds will likely require observing black holes before they have grown orders of magnitude in mass. Indeed, it has been proposed that DCBH formation leads to so-called ``obese black hole galaxies'' which have black hole masses comparable to their stellar masses \citep{2013MNRAS.432.3438A, 2017ApJ...838..117N}. 

Here we further explore the possibility of distinguishing high-redshift DCBH seeds due to their large initial black hole mass to halo mass ratios. We utilize cosmological N-body simulations to follow a large sample of ACHs (sites of potential DCBH formation) and demonstrate that, before they grow more than roughly an order of magnitude, essentially all DCBHs will be hosted by halos or subhalos with black hole mass to halo mass ratios much higher than would be expected from Pop III seeds which have grown to the same mass.
Thus, before they grow in mass
significantly, DCBHs can be observed in X-rays with future instruments such as \emph{Lynx} \citep{2017SPIE10397E..0SG}, and will be distinguishable from Pop III seeds due to their small stellar mass (a consequence of their small halo masses), which can be measured with infrared telescopes such as the \emph{James Webb Space Telescope} \citep[\emph{JWST}; see related discussion in section 4.1 of][]{2018MNRAS.479.4017I}. We note that our test also applies to a Pop III seed which grows rapidly via hyper-Eddington accretion into a ${\sim}10^5~M_\odot$ black hole seed in an ACH \citep{2016MNRAS.459.3738I, 2016MNRAS.460.4122R}, which would be indistinguishable from a DCBH in this context.

This Letter is structured as follows. In Section 2 we describe the cosmological simulation used to analyze the evolution of ACHs. We describe our results in Section 3 and discuss our main conclusions in Section 4. Throughout, we assume a $\Lambda {\rm CDM}$ cosmology with parameters consistent with those reported by the \cite{2016A&A...596A.108P}: $\Omega_{\rm m}=0.32$, $\Omega_{\Lambda}=0.68$, $\Omega_{\rm b}=0.049$, $h=0.67$, $\sigma_8=0.83$, and $n_{\rm s}=0.96$.

\section{Simulation}
We have performed a cosmological N-body simulation with the publicly available code \textsc{gadget2} \citep{2001NewA....6...79S} utilizing initial conditions generated by \textsc{2lptic} \citep{2006MNRAS.373..369C}. Our simulation has a resolution of $1024^3$ particles and a box size of 20 comoving Mpc. The simulation resolves $3\times 10^7~ M_\odot$ ACHs with ${\sim} 100$ particles each, which is sufficient to track these halos as substructures of larger systems \citep{2012MNRAS.423.1200O}. We save simulation snapshots at $15~{\rm Myr}$ intervals from $z=10$ to $z=6$ and use the \textsc{rockstar} \citep{2013ApJ...762..109B} halo finder with \textsc{consistent trees} \citep{2013ApJ...763...18B} to construct merger trees. 

\section{Results}
\subsection{BH Mass to Stellar Mass Ratio}
We track the evolution of ACHs in our simulation to demonstrate that before they grow significantly in mass (within $\sim 75~{\rm Myr}$ of their birth), newly formed DCBHs remain in halos (or subhalos) with black hole mass to stellar mass ratios well above what is expected in the Pop III seed scenario. Note that most ACHs are not thought to host DCBH formation, but as described below, our conclusions apply to  ${\sim}10^5~M_\odot$ black hole seeds formed in almost any ACH. We identify the $7929$ halos which cross the atomic cooling threshold ($M_{\rm a} = 3 \times 10^7 ~M_\odot$) between $z=10$ and the next simulation snapshot 15 Myr later ($z=9.8$). We then follow the properties of these halos through time. For those which fall into a larger halo but remain a subhalo, we record both the subhalo and host halo properties.

Assuming Eddington-limited growth as expected for black holes with $M_{\rm BH}\gtrsim10^5~M_\odot$ \citep{2016MNRAS.459.3738I}, we estimate the mass of a growing DCBH as $M_{\rm BH} = M_{\rm i} e^{ t / t_{\rm E}}$, where $t_{\rm E}$ is the $e$-folding timescale set by the Eddington limit, $M_{\rm i}$ is the initial seed mass, and $t$ is the duration of time since DCBH formation (assumed to coincide with the second snapshot in our simulation just after ACH formation). We then compute the black hole mass to stellar mass of each ACH by assuming 
the stellar mass is given by $M_* = M_{\rm h} \frac{\Omega_b}{\Omega_{m}} f_*$, where $M_{\rm h}$ is the total halo mass and $f_*$ is the star formation efficiency. 
We use fiducial values of $M_{\rm i} = 10^5~M_\odot$, $t_E = 30~{\rm Myr}$ and $f_*=0.05$ to estimate the black hole mass to stellar mass ratio  for a DCBH in each ACH as a function of cosmic time. We report this ratio as  $\frac{M_{\rm BH}}{M_*}\left(\frac{0.05}{f_*} \right )$, due to its simple dependence on the star formation efficiency. Our assumption of $t_E = 30~{\rm Myr}$ corresponds to a radiative efficiency of $\epsilon \approx 0.06$ which is predicted for a Schwarzschild black hole. Our conclusions do not depend sensitively on this exact choice. We note that the environments of newly formed DCBHs may lead to outflows driven by radiative pressure that could reduce $f_*$ below our fiducial value \citep{2017MNRAS.472..205S}. This would serve to increase $\frac{M_{\rm BH}}{M_*}$, strengthening our overall conclusions.

\begin{figure*}
\includegraphics[clip=false,keepaspectratio=true, width = 90mm]{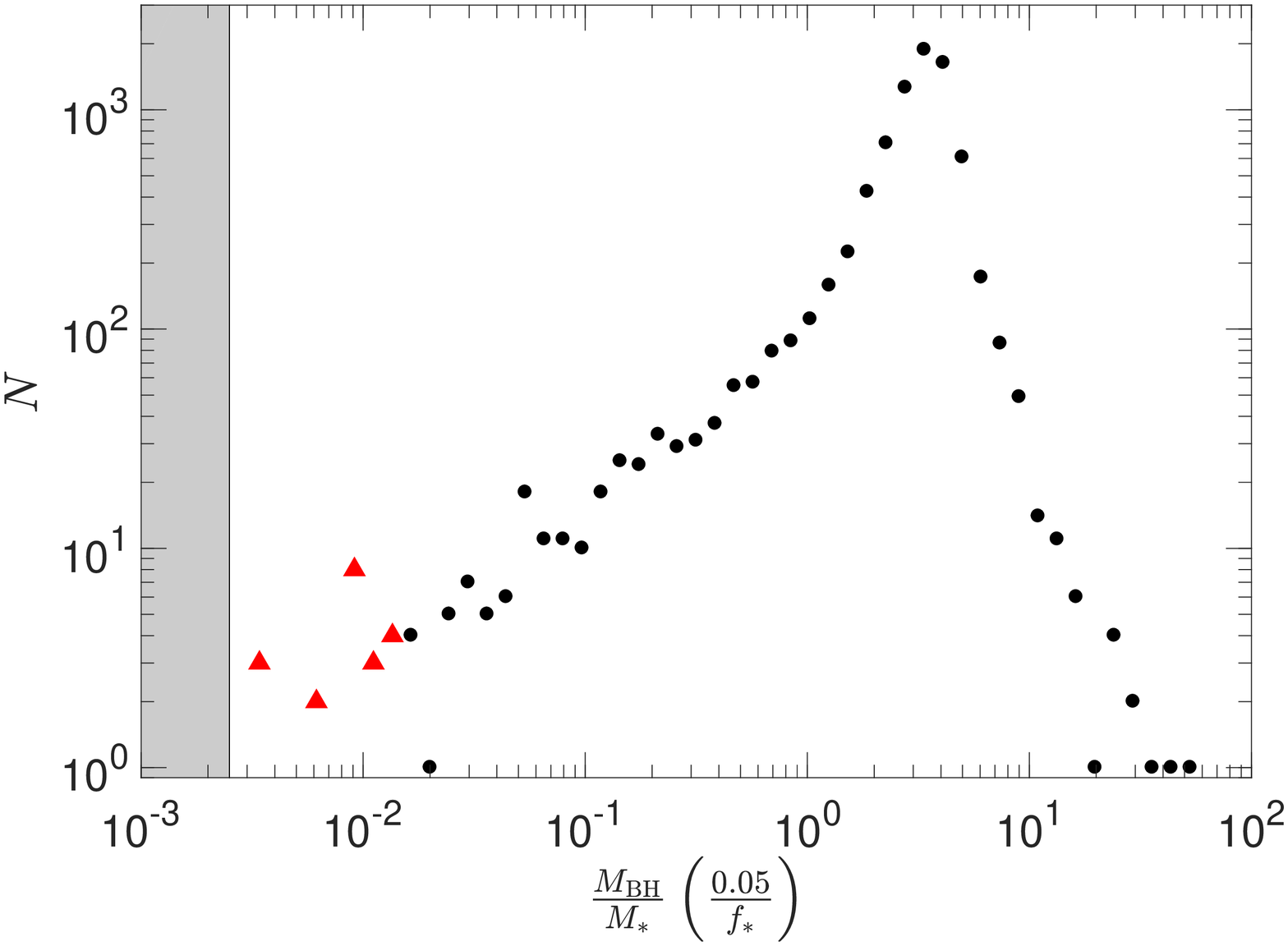}
\includegraphics[clip=false,keepaspectratio=true, width = 90mm]{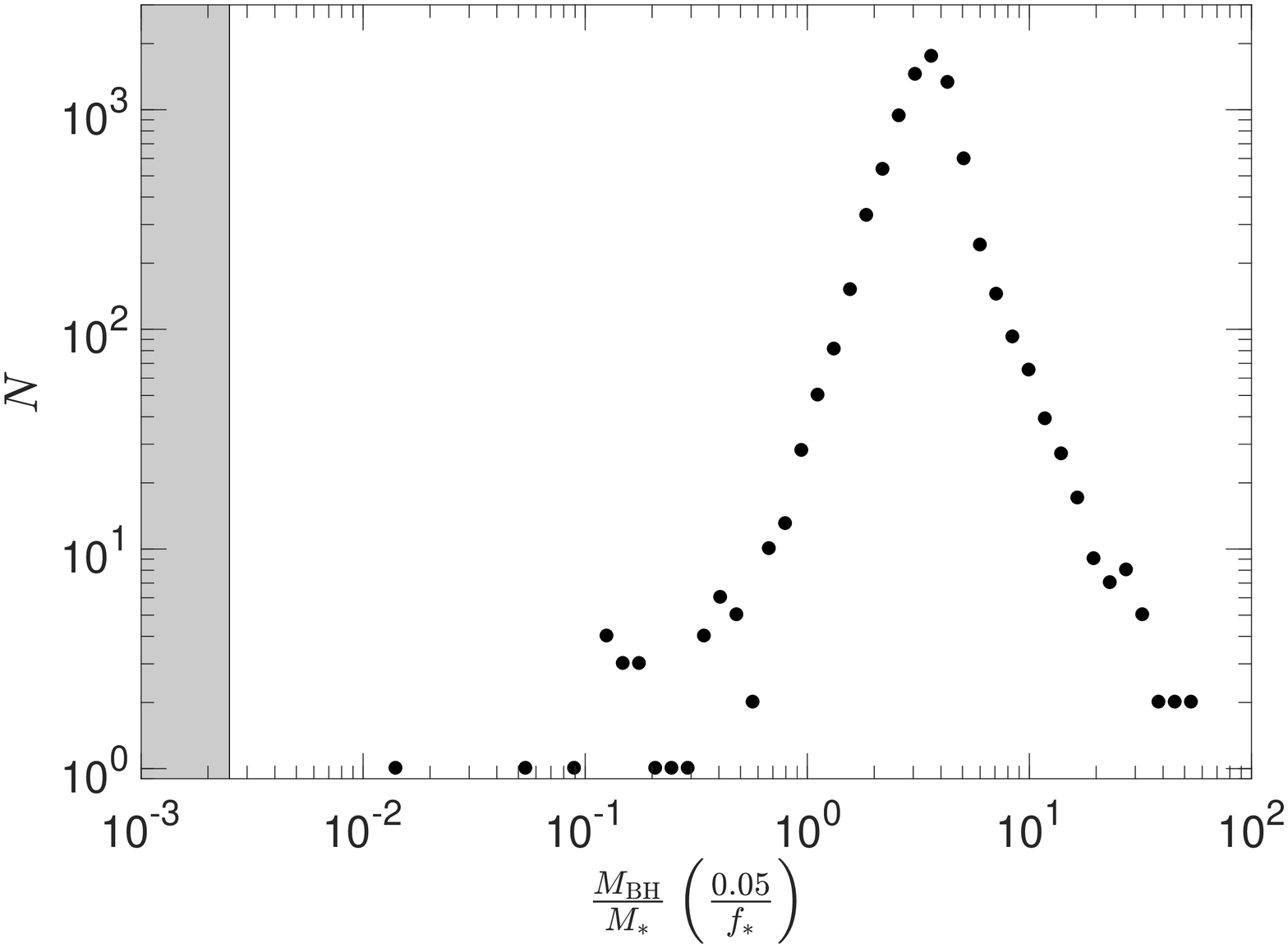}
\caption{\label{hist}  The distribution of black hole mass to stellar mass ratios for halos hosting growing DCBHs 75 Myr after their formation ($z=8.8$). The left panel uses the stellar mass for the largest halo hosting the DCBH (i.e.~if it is in a subhalo, the larger host halo mass is used), while the right panel uses the stellar mass of the subhalo when a DCBH resides in a satellite. For reference, we have plotted the upper limit for Pop III seeds simulated by \cite{2017MNRAS.468.3935H} (right edges of shaded grey regions).  The triangles in the left panel denote ACHs which merge with halos that have masses greater than $10^{10}~M_\odot$. DCBHs in these halos almost always orbit as satellites which can be resolved with future X-ray and infrared observations (see Figure \ref{dist}) and correspond to $\frac{M_{\rm BH}}{M_*}\left(\frac{0.05}{f_*} \right ) \approx 5$ in the right panel. } 
\end{figure*}

In Figure \ref{hist}, we plot the distribution of $\frac{M_{\rm BH}}{M_*}\left(\frac{0.05}{f_*} \right )$ at $z=8.8$, 75 Myr after the formation of our ACHs. By this time the DCBHs are assumed to have grown to $M_{\rm BH}=1.2\times10^{6}~M_\odot$. We compare this to the black hole mass to stellar mass ratio expected for ``light'' Pop III seeds. Although their initial mass function remains uncertain, the first Pop III stars are thought to have masses of ${\sim} 10-1000~M_\odot$ \citep{2015MNRAS.448..568H} and form at very high redshift in small ${\sim}10^{5-6}~M_\odot$ dark matter ``minihalos''.

The growth of Pop III black hole remnants has been tracked in simulations \citep{2009ApJ...701L.133A, 2014MNRAS.442.2751T, 2017MNRAS.468.3935H, 2018arXiv180406477S} and semi-analytic calculations \citep{2003ApJ...582..559V, 2009ApJ...696.1798T}.
To make our comparisons we focus on the cosmological hydrodynamical simulations of \cite{2017MNRAS.468.3935H}. These simulations seed SMBHs on the fly in dense low-metallicity gas with ${\sim}1000~M_\odot$ black holes \citep[in the same vein as][]{2011ApJ...742...13B}, which then grow with an accretion rate capped by the Eddington limit. Several different supernovae feedback prescriptions are utilized. For the weaker thermal feedback prescription, the largest ratio found for all of the galaxies with $M_{\rm BH}\approx 10^6~M_\odot$ is $M_{\rm BH}/M_* \approx 2.5\times 10^{-3}$, which we plot for comparison with our DCBHs. This is a conservative choice, as their stronger (more realistic) feedback prescriptions give lower values of this ratio. We also note that the semi-analytic simulations of \cite{2003ApJ...582..559V} (see their Figure 11) find $\frac{M_{\rm BH}}{M_*}\left(\frac{0.05}{f_*} \right )  \approx 10^{-5}$ at $M_{\rm BH} \approx 10^6~M_\odot$ for two representative cases, which is signicantly lower than our conservative limit in Figure \ref{hist} (see also \cite{2018arXiv180406477S}, who find even less Pop III seed growth). 
Additionally, inefficient accretion due to radiative feedback from the growing black hole itself \citep{2009ApJ...696L.146M} would lower $M_{\rm BH}/M_*$ in the Pop III case.

In the left panel of Figure \ref{hist}, $\frac{M_{\rm BH}}{M_*}\left(\frac{0.05}{f_*} \right )$ is plotted for the largest halo hosting the DCBHs (i.e.~if it is in a subhalo, we use the larger host halo mass to compute $M_*$). While the vast majority of ACHs have several orders of magnitude higher $M_{\rm BH}/M_*$ than the Pop III limit, we do find a small fraction of halos relatively close to this value. This is because some ACHs quickly fall into a much larger nearby halo. Fortunately, almost all of these ACHs remain subhalos separated from their larger host halos. This can be seen in the right panel of Figure \ref{hist}, where we show $\frac{M_{\rm BH}}{M_*}\left(\frac{0.05}{f_*} \right )$  computing the stellar mass from the subhalo mass if the DCBH is in a substructure within the larger halo. We note that when an ACH falls into massive neighbor ($M_{\rm h} > 10^{10}~M_\odot$), its total mass never exceeds the atomic cooling threshold by more than a factor of two. Except for one ACH which has completely merged and mixed with its larger neighbor (the point with the lowest ratio in the right panel of Figure \ref{hist}), the black hole mass to stellar mass ratio for our DCBHs is more than an order of magnitude higher than our conservative upper limit in the Pop III scenario. 

\begin{figure}
\plotone{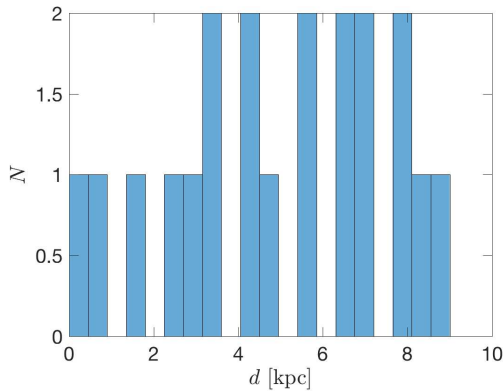}
\caption{\label{dist}  The distribution of separations (in physical units) between our DCBH-containing subhalos and the larger host halos for all 20 cases with  $\frac{M_{\rm BH}}{M_*}\left(\frac{0.05}{f_*} \right ) < 0.015$ in the left panel of Figure \ref{hist} (red triangles). A separation of 1 kpc corresponds to $0.22''$ at $z=8.8$. Almost all DCBHs will be separated by distances large enough to be spatially resolved by the future X-ray telescope \emph{Lynx} (${\sim}0.2''$ positional accuracy) and with \emph{JWST} (${\sim}0.1''$ angular resolution) .} 
\end{figure}

Importantly, we find that ACHs which become satellites of large neighbors remain separated from their larger host halos by distances which can be resolved with future observations. In Figure \ref{dist}, we plot the distances between the centers of DCBH-hosting subhalos and their larger host halos for the 20 ACHs that end up in host halos with $M_{\rm h} > 10^{10}~M_\odot$ (corresponding to $\frac{M_{\rm BH}}{M_*}\left(\frac{0.05}{f_*} \right ) < 0.015$). Besides one ACH which has completely merged, these satellite halos are at separations $\gtrsim 1~{\rm kpc}$. A physical distance of 1 kpc corresponds to $0.22''$ at $z=8.8$. Thus, almost all of these halos are sufficiently separated to be spatially resolved in X-ray observations with \emph{Lynx}, which is planned to have $0.5''$ angular resolution and $0.2''$ positional accuracy for faint sources (Alexey Vikhlinin, private communication), and by infrared observations with \emph{JWST} (${\sim}0.1''$ resolution). Thus, from Figures \ref{hist} and \ref{dist}, we can see that except for the one ACH which completely merges and mixes with a massive neighbor, DCBHs seeds have $\frac{M_{\rm BH}}{M_*}\left(\frac{0.05}{f_*} \right ) \gtrsim0.1$, even after they have grown an order of magnitude in mass. Thus we propose that identifying DCBHs with X-ray observations and comparing with infrared observations to put limits on the stellar mass can likely distinguish DCBHs from Pop III seeds. 
In the DCBH case, we expect \emph{JWST} to detect a very small stellar host (or none at all), or else detect a galaxy that is offset by ${\sim} 1'' $ from the black hole X-ray source. In the Pop III case, we expect JWST to detect a much brighter host in the near infrared with $M_{\rm BH}/M_* < 2.5\times 10^{-3}$.

\subsection{Satellite Halos}
As discussed above, we find one ACH (out of $\sim$8000) which completely merges and mixes with a very massive ($M_{\rm h}>10^{10}~M_\odot$) neighbor within 75 Myr after DCBH formation. A DCBH in such a halo would be more challenging to distinguish from a black hole grown from a Pop III seed due to its lower black hole mass to stellar mass ratio. Because the number density of ${\sim}10^9~M_\odot$ SMBHs at $z\gtrsim6$ is very low (${\sim} 1~{\rm Gpc}^{-3}$), it is important to check that it is not halos like this 1 in ${\sim}8000$ ACH which are biased towards DCBH formation. It is generally thought that DCBHs form preferentially in ACHs exposed to strong LW radiation, so one might worry that it is the halos with the highest LW exposure that end up merging completely with massive neighbors, making them hard to distinguish from Pop III seeds.
Here we explain that the complete merger is mainly due to the orbital properties of the ACH and larger neighbor, and that DCBH formation should not be biased to occur in a halo like this compared to other ACHs which merge with their massive neighbor but remain in satellites which can be spatially resolved.

\begin{figure*}
\plottwo{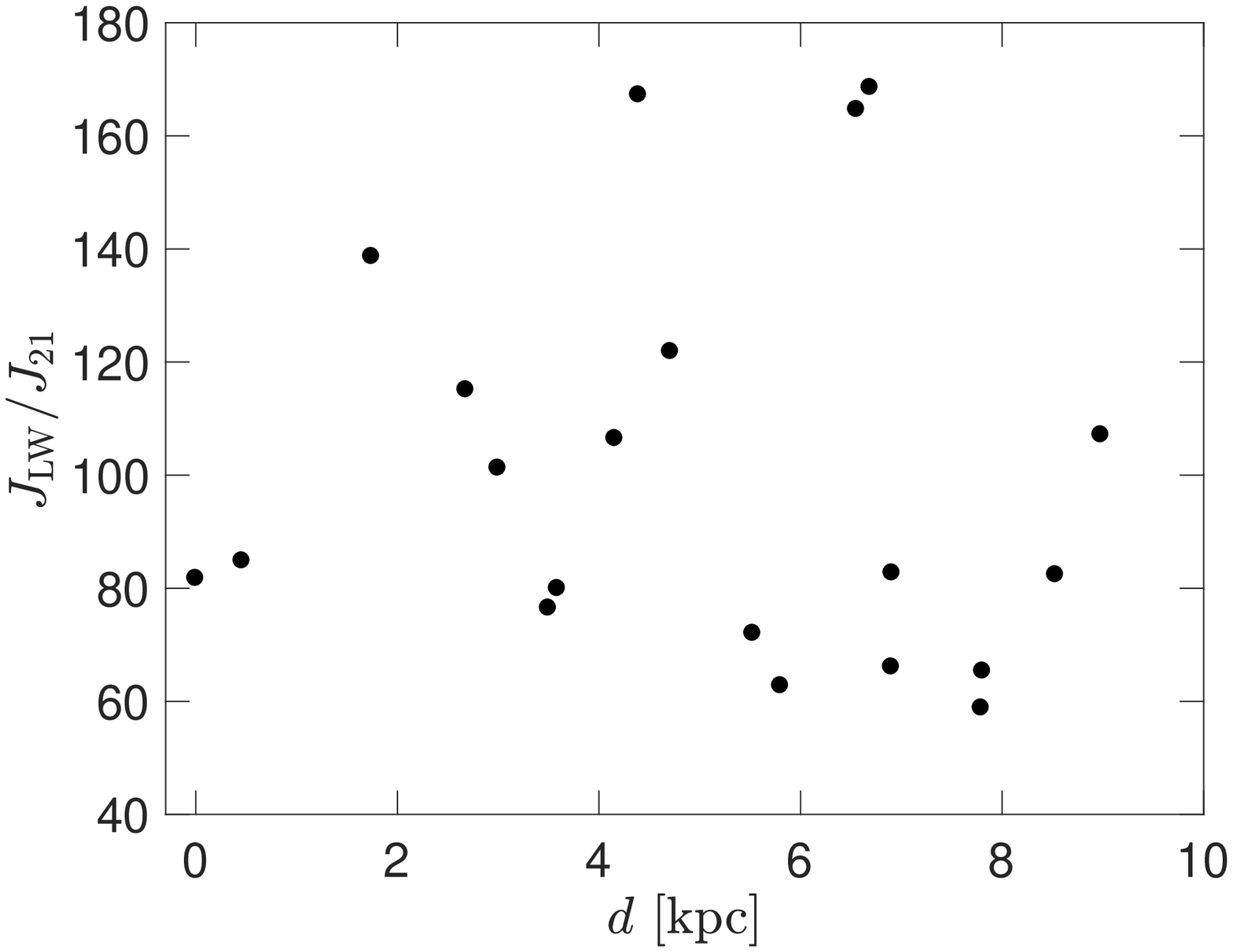}{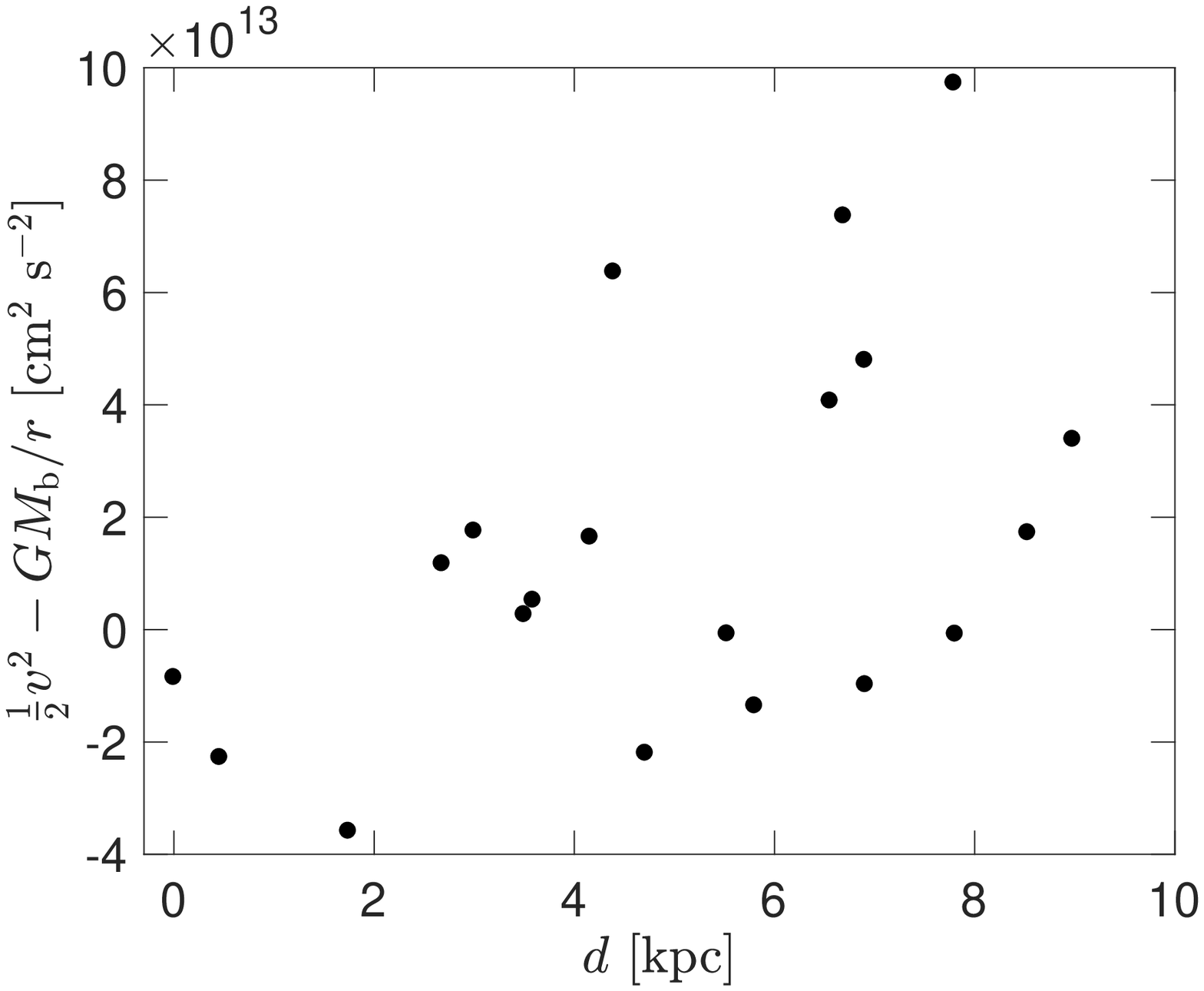}
\caption{\label{scatter} The LW intensity normalized by $J_{21}=10^{-21}~{\rm erg~s^{-1}~cm^{-2}~Hz^{-1}~sr^{-1}}$ (left panel) and $\frac{1}{2} v^2 - GM_{\rm b}/r$ (right panel) at the time of DCBH formation ($z=9.8$) for all ACHs from Figure \ref{dist} versus the distance from the center of the massive neighbor 75 Myr after DCBH formation (at $z=8.8$). 
The LW flux at $z=9.8$ has very little correlation with the separation at $z=8.8$, but there is a correlation for $\frac{1}{2} v^2 - GM_{\rm b}/r$, with more bound halos ending up at smaller separations.  The ACH which completely merges and mixes with a massive neighbor halo (i.e.~$d=0$) has low $J_{\rm LW}$ and is more tightly bound than most. We conclude that DCBH formation should not be biased towards ACHs which completely merge and mix with massive neighbors. }
\end{figure*}

In Figure \ref{scatter}, we plot properties of the 20 ACHs (at the time of DCBH formation) that eventually merge with massive neighbors ($M_{\rm h}>10^{10}~M_\odot$). In the right panel, we plot the Lyman-Werner (LW) intensity seen by each ACH at the time of DCBH formation as a function of the distance from the host halo 75 Myr later (i.e.~the LW intensity at $z=9.8$ and the distance at $z=8.8$). The LW intensity is computed by adding the contribution from all halos in the simulation box above the atomic cooling threshold (halos seeing the strongest LW flux have a negligible contribution from the background outside of the box, most of the flux comes from nearby neighbors). Each halo's LW luminosity is computed by assuming that 4000 LW photons are produced per baryon incorporated into stars and that the star formation rate is given by ${\rm SFR}= M_*/(0.1t_{\rm H})$, where $M_* = M_{\rm h} \frac{\Omega_b}{\Omega_{m}} f_*$ and $0.1t_{\rm H}$ is the dynamical time of the halo at that redshift.  From the left panel of Figure \ref{scatter}, we see that, among the ACHs which merge with large halos, there is no clear correlation between their initial observed LW intensity and the distance they end up as satellites from the larger halo after merger. Thus, even if DCBHs form preferentially near massive neighbors, the vast majority, $\sim 90 \%$ (18 out of 20 halos in our simulation), are likely to remain in satellite halos at distances from their host which can be spatially resolved with \emph{Lynx}.

So what causes one halo to completely merge and mix into a massive neighbor, but not the others? We find that the distance between the satellites and their host halos at $z=8.8$ correlates with how bound their orbits are at DCBH formation ($z=9.8$). This is illustrated in the right panel of Figure \ref{scatter} where we show $\frac{1}{2} v^2 - GM_{\rm b}/r$ at the time of DCBH formation versus the separation 75 Myr later. Here $v$ is the velocity of the ACH relative to the massive neighbor, $M_{\rm b}$ is the mass of the large neighbor, and $r$ is their separation at $z=9.8$. More tightly bound halos (lower values of this quantity) systematically end up at smaller separations from their massive neighbors. The halo which completely merges (i.e.~$d=0$) is one of the more tightly bound halos, but not the most tightly bound. Its complete merger is likely determined by the detailed characteristics of its orbit. There is no obvious reason DCBH formation should preferentially occur in ACHs which are tightly bound and on orbits which will mix with larger neighbors. Thus, we conclude that within 75 Myr, the vast majority of DCBHs will either be in a halo or subhalo sufficiently small such that the high black hole mass to stellar mass ratio will make the heavy seed distinguishable from Pop III models.

\section{Conclusions}
While DCBH formation has been proposed to explain the rapid growth of the first SMBHs in the Universe, we still lack observational evidence confirming or refuting this picture. Some promising tests which have been proposed include examining SBMHs spectral energy distributions \citep{2015MNRAS.454.3771P, 2016MNRAS.459.1432P}, looking for fossil evidence from local black holes in low mass galaxies \citep{2008MNRAS.383.1079V, 2012NatCo...3E1304G}, and searching for obese black hole galaxies \citep{2013MNRAS.432.3438A, 2017ApJ...838..117N}.  In this Letter, building on the idea of obese black hole galaxies, we demonstrate that before they grow more than an order of magnitude in mass, DCBHs should be distinguishable from Pop III black hole seeds with feedback-limited growth due to their very high black hole mass to stellar mass ratios. Utilizing a cosmological N-body simulation to track the evolution of ${\sim} 8000$ ACHs (the potential birth sites of DCBHs) and making simple assumptions regarding black hole growth and star formation, we find that essentially all DCBHs have $\frac{M_{\rm BH}}{M_*}\left(\frac{0.05}{f_*} \right )  > 0.1$. Conservatively, this is more than an order of magnitude higher than expected in the Pop III seed scenario. 

We find 20 ACHs which fall into nearby halos with $M_{\rm h}>10^{10}~M_\odot$, however in almost all cases the ACHs orbit the larger halo as subhalos with separations that can be spatially resolved with future X-ray and infrared observations. Small separations correlate with more bound orbits and the rare case (1 out of ${\sim}8000$) of a merger which completely mixes into one distinct halo is due to the detailed properties of the orbit which are not expected to strongly correlate with DCBH formation. 

We have shown that it is possible to distinguish if a  ${\sim}10^6~M_\odot$ black hole has grown from a DCBH or a Pop III seed by observing the black hole mass to stellar mass ratio. We propose that future X-ray observations should search for accreting black holes with $M_{\rm BH} \approx 10^6~M_\odot$ at $z \approx 10$. Follow-up or precursor infrared observations should then be compared and if a limit of $M_*  \lesssim 5\times 10^8~M_\odot$ is observed at the location of the black hole (though note that if it merged with a larger halo it may be spatially offset from a separate galaxy), it would be strong evidence for DCBH formation. Future telescopes such as \emph{Lynx} and  \emph{JWST} should have sufficient sensitivity and angular resolution to achieve this.

Finally, we note that we have assumed Eddington-limited accretion throughout our analysis. This is theoretically motived for $M_{\rm BH} \gtrsim 10^5~M_\odot$ \citep{2016MNRAS.459.3738I}, however it may be possible for a light seed to rapidly grow to ${\sim}10^5~M_\odot$ via hyper-Eddington accretion in an ACH \citep{2016MNRAS.459.3738I, 2016MNRAS.460.4122R}. This would be indistinguishable from a DCBH via the test proposed here. Thus, we emphasize that, strictly speaking, the black hole mass to stellar mass ratio discriminates between heavy seeds formed in ACHs versus Pop III seeds with feedback-limited accretion.

\section*{Acknowledgements}
We thank Alexey Vikhlinin and Niel Brandt for useful discussions. We also thank Greg Bryan, Melanie Habouzit, Kohei Inayoshi, and Fabio Pacucci for helpful comments. The Flatiron Institute is supported by the Simons Foundation. The simulations were carried out on the Flatiron Institute supercomputer, Rusty. ZH was supported in part by NASA grant NNX15AB19G. 

\bibliography{DCBH_letter}

\begin{thebibliography}{}
\expandafter\ifx\csname natexlab\endcsname\relax\def\natexlab#1{#1}\fi
\providecommand{\url}[1]{\href{#1}{#1}}
\providecommand{\dodoi}[1]{doi:~\href{http://doi.org/#1}{\nolinkurl{#1}}}
\providecommand{\doeprint}[1]{\href{http://ascl.net/#1}{\nolinkurl{http://ascl.net/#1}}}
\providecommand{\doarXiv}[1]{\href{https://arxiv.org/abs/#1}{\nolinkurl{https://arxiv.org/abs/#1}}}

\bibitem[{{Agarwal} {et~al.}(2013){Agarwal}, {Davis}, {Khochfar}, {Natarajan},
  \& {Dunlop}}]{2013MNRAS.432.3438A}
{Agarwal}, B., {Davis}, A.~J., {Khochfar}, S., {Natarajan}, P., \& {Dunlop},
  J.~S. 2013, \mnras, 432, 3438, \dodoi{10.1093/mnras/stt696}

\bibitem[{{Alvarez} {et~al.}(2009){Alvarez}, {Wise}, \&
  {Abel}}]{2009ApJ...701L.133A}
{Alvarez}, M.~A., {Wise}, J.~H., \& {Abel}, T. 2009, \apjl, 701, L133,
  \dodoi{10.1088/0004-637X/701/2/L133}

\bibitem[{{Behroozi} {et~al.}(2013{\natexlab{a}}){Behroozi}, {Wechsler}, \&
  {Wu}}]{2013ApJ...762..109B}
{Behroozi}, P.~S., {Wechsler}, R.~H., \& {Wu}, H.-Y. 2013{\natexlab{a}}, \apj,
  762, 109, \dodoi{10.1088/0004-637X/762/2/109}

\bibitem[{{Behroozi} {et~al.}(2013{\natexlab{b}}){Behroozi}, {Wechsler}, {Wu},
  {Busha}, {Klypin}, \& {Primack}}]{2013ApJ...763...18B}
{Behroozi}, P.~S., {Wechsler}, R.~H., {Wu}, H.-Y., {et~al.} 2013{\natexlab{b}},
  \apj, 763, 18, \dodoi{10.1088/0004-637X/763/1/18}

\bibitem[{{Bellovary} {et~al.}(2011){Bellovary}, {Volonteri}, {Governato},
  {Shen}, {Quinn}, \& {Wadsley}}]{2011ApJ...742...13B}
{Bellovary}, J., {Volonteri}, M., {Governato}, F., {et~al.} 2011, \apj, 742,
  13, \dodoi{10.1088/0004-637X/742/1/13}

\bibitem[{{Bromm} \& {Loeb}(2003)}]{2003ApJ...596...34B}
{Bromm}, V., \& {Loeb}, A. 2003, \apj, 596, 34, \dodoi{10.1086/377529}

\bibitem[{{Crocce} {et~al.}(2006){Crocce}, {Pueblas}, \&
  {Scoccimarro}}]{2006MNRAS.373..369C}
{Crocce}, M., {Pueblas}, S., \& {Scoccimarro}, R. 2006, \mnras, 373, 369,
  \dodoi{10.1111/j.1365-2966.2006.11040.x}

\bibitem[{{Di Matteo} {et~al.}(2008){Di Matteo}, {Colberg}, {Springel},
  {Hernquist}, \& {Sijacki}}]{2008ApJ...676...33D}
{Di Matteo}, T., {Colberg}, J., {Springel}, V., {Hernquist}, L., \& {Sijacki},
  D. 2008, \apj, 676, 33, \dodoi{10.1086/524921}

\bibitem[{{Dijkstra} {et~al.}(2014){Dijkstra}, {Ferrara}, \&
  {Mesinger}}]{2014MNRAS.442.2036D}
{Dijkstra}, M., {Ferrara}, A., \& {Mesinger}, A. 2014, \mnras, 442, 2036,
  \dodoi{10.1093/mnras/stu1007}

\bibitem[{{Dijkstra} {et~al.}(2008){Dijkstra}, {Haiman}, {Mesinger}, \&
  {Wyithe}}]{2008MNRAS.391.1961D}
{Dijkstra}, M., {Haiman}, Z., {Mesinger}, A., \& {Wyithe}, J.~S.~B. 2008,
  \mnras, 391, 1961, \dodoi{10.1111/j.1365-2966.2008.14031.x}

\bibitem[{{Eisenstein} \& {Loeb}(1995)}]{1995ApJ...443...11E}
{Eisenstein}, D.~J., \& {Loeb}, A. 1995, \apj, 443, 11, \dodoi{10.1086/175498}

\bibitem[{{Gaskin} {et~al.}(2017){Gaskin}, {Allured}, {Bandler}, {Basso},
  {Bautz}, {Baysinger}, {Biskach}, {Boswell}, {Capizzo}, {Chan}, {Civitani},
  {Cohen}, {Cotroneo}, {Davis}, {DeRoo}, {DiPirro}, {Dominguez}, {Fabisinski},
  {Falcone}, {Figueroa-Feliciano}, {Garcia}, {Gelmis}, {Heilmann}, {Hopkins},
  {Jackson}, {Kilaru}, {Kraft}, {Liu}, {McClelland}, {McEntaffer}, {McCarley},
  {Mulqueen}, {{\"O}zel}, {Pareschi}, {Reid}, {Riveros}, {Rodriguez}, {Rowe},
  {Saha}, {Schattenburg}, {Schnell}, {Schwartz}, {Solly}, {Suggs}, {Sutherlin},
  {Swartz}, {Trolier-McKinstry}, {Tutt}, {Vikhlinin}, {Walker}, {Yoon}, \&
  {Zhang}}]{2017SPIE10397E..0SG}
{Gaskin}, J.~A., {Allured}, R., {Bandler}, S.~R., {et~al.} 2017, in Society of
  Photo-Optical Instrumentation Engineers (SPIE) Conference Series, Vol. 10397,
  Society of Photo-Optical Instrumentation Engineers (SPIE) Conference Series,
  103970S

\bibitem[{{Greene}(2012)}]{2012NatCo...3E1304G}
{Greene}, J.~E. 2012, Nature Communications, 3, 1304,
  \dodoi{10.1038/ncomms2314}

\bibitem[{{Habouzit} {et~al.}(2017){Habouzit}, {Volonteri}, \&
  {Dubois}}]{2017MNRAS.468.3935H}
{Habouzit}, M., {Volonteri}, M., \& {Dubois}, Y. 2017, \mnras, 468, 3935,
  \dodoi{10.1093/mnras/stx666}

\bibitem[{{Haiman}(2013)}]{2013ASSL..396..293H}
{Haiman}, Z. 2013, in Astrophysics and Space Science Library, Vol. 396, The
  First Galaxies, ed. T.~{Wiklind}, B.~{Mobasher}, \& V.~{Bromm}, 293

\bibitem[{{Hirano} {et~al.}(2017){Hirano}, {Hosokawa}, {Yoshida}, \&
  {Kuiper}}]{2017Sci...357.1375H}
{Hirano}, S., {Hosokawa}, T., {Yoshida}, N., \& {Kuiper}, R. 2017, Science,
  357, 1375, \dodoi{10.1126/science.aai9119}

\bibitem[{{Hirano} {et~al.}(2015){Hirano}, {Hosokawa}, {Yoshida}, {Omukai}, \&
  {Yorke}}]{2015MNRAS.448..568H}
{Hirano}, S., {Hosokawa}, T., {Yoshida}, N., {Omukai}, K., \& {Yorke}, H.~W.
  2015, \mnras, 448, 568, \dodoi{10.1093/mnras/stv044}

\bibitem[{{Inayoshi} {et~al.}(2016){Inayoshi}, {Haiman}, \&
  {Ostriker}}]{2016MNRAS.459.3738I}
{Inayoshi}, K., {Haiman}, Z., \& {Ostriker}, J.~P. 2016, \mnras, 459, 3738,
  \dodoi{10.1093/mnras/stw836}

\bibitem[{{Inayoshi} {et~al.}(2018){Inayoshi}, {Li}, \&
  {Haiman}}]{2018MNRAS.479.4017I}
{Inayoshi}, K., {Li}, M., \& {Haiman}, Z. 2018, \mnras, 479, 4017,
  \dodoi{10.1093/mnras/sty1720}

\bibitem[{{Inayoshi} {et~al.}(2015){Inayoshi}, {Visbal}, \&
  {Kashiyama}}]{2015MNRAS.453.1692I}
{Inayoshi}, K., {Visbal}, E., \& {Kashiyama}, K. 2015, \mnras, 453, 1692,
  \dodoi{10.1093/mnras/stv1654}

\bibitem[{{Lodato} \& {Natarajan}(2006)}]{2006MNRAS.371.1813L}
{Lodato}, G., \& {Natarajan}, P. 2006, \mnras, 371, 1813,
  \dodoi{10.1111/j.1365-2966.2006.10801.x}

\bibitem[{{Milosavljevi{\'c}} {et~al.}(2009){Milosavljevi{\'c}}, {Couch}, \&
  {Bromm}}]{2009ApJ...696L.146M}
{Milosavljevi{\'c}}, M., {Couch}, S.~M., \& {Bromm}, V. 2009, \apjl, 696, L146,
  \dodoi{10.1088/0004-637X/696/2/L146}

\bibitem[{{Mortlock} {et~al.}(2011){Mortlock}, {Warren}, {Venemans}, {Patel},
  {Hewett}, {McMahon}, {Simpson}, {Theuns}, {Gonz{\'a}les-Solares}, {Adamson},
  {Dye}, {Hambly}, {Hirst}, {Irwin}, {Kuiper}, {Lawrence}, \&
  {R{\"o}ttgering}}]{2011Natur.474..616M}
{Mortlock}, D.~J., {Warren}, S.~J., {Venemans}, B.~P., {et~al.} 2011, \nat,
  474, 616, \dodoi{10.1038/nature10159}

\bibitem[{{Natarajan} {et~al.}(2017){Natarajan}, {Pacucci}, {Ferrara},
  {Agarwal}, {Ricarte}, {Zackrisson}, \& {Cappelluti}}]{2017ApJ...838..117N}
{Natarajan}, P., {Pacucci}, F., {Ferrara}, A., {et~al.} 2017, \apj, 838, 117,
  \dodoi{10.3847/1538-4357/aa6330}

\bibitem[{{Oh} \& {Haiman}(2002)}]{2002ApJ...569..558O}
{Oh}, S.~P., \& {Haiman}, Z. 2002, \apj, 569, 558, \dodoi{10.1086/339393}

\bibitem[{{Omukai}(2001)}]{2001ApJ...546..635O}
{Omukai}, K. 2001, \apj, 546, 635, \dodoi{10.1086/318296}

\bibitem[{{Onions} {et~al.}(2012){Onions}, {Knebe}, {Pearce}, {Muldrew}, {Lux},
  {Knollmann}, {Ascasibar}, {Behroozi}, {Elahi}, {Han}, {Maciejewski},
  {Merch{\'a}n}, {Neyrinck}, {Ruiz}, {Sgr{\'o}}, {Springel}, \&
  {Tweed}}]{2012MNRAS.423.1200O}
{Onions}, J., {Knebe}, A., {Pearce}, F.~R., {et~al.} 2012, \mnras, 423, 1200,
  \dodoi{10.1111/j.1365-2966.2012.20947.x}

\bibitem[{{Pacucci} {et~al.}(2016){Pacucci}, {Ferrara}, {Grazian}, {Fiore},
  {Giallongo}, \& {Puccetti}}]{2016MNRAS.459.1432P}
{Pacucci}, F., {Ferrara}, A., {Grazian}, A., {et~al.} 2016, \mnras, 459, 1432,
  \dodoi{10.1093/mnras/stw725}

\bibitem[{{Pacucci} {et~al.}(2015){Pacucci}, {Ferrara}, {Volonteri}, \&
  {Dubus}}]{2015MNRAS.454.3771P}
{Pacucci}, F., {Ferrara}, A., {Volonteri}, M., \& {Dubus}, G. 2015, \mnras,
  454, 3771, \dodoi{10.1093/mnras/stv2196}

\bibitem[{{Pacucci} {et~al.}(2017){Pacucci}, {Natarajan}, {Volonteri},
  {Cappelluti}, \& {Urry}}]{2017ApJ...850L..42P}
{Pacucci}, F., {Natarajan}, P., {Volonteri}, M., {Cappelluti}, N., \& {Urry},
  C.~M. 2017, \apjl, 850, L42, \dodoi{10.3847/2041-8213/aa9aea}

\bibitem[{{Planck Collaboration} {et~al.}(2016){Planck Collaboration}, {Adam},
  {Aghanim}, {Ashdown}, {Aumont}, {Baccigalupi}, {Ballardini}, {Banday},
  {Barreiro}, {Bartolo}, {Basak}, {Battye}, {Benabed}, {Bernard}, {Bersanelli},
  {Bielewicz}, {Bock}, {Bonaldi}, {Bonavera}, {Bond}, {Borrill}, {Bouchet},
  {Boulanger}, {Bucher}, {Burigana}, {Calabrese}, {Cardoso}, {Carron},
  {Chiang}, {Colombo}, {Combet}, {Comis}, {Couchot}, {Coulais}, {Crill},
  {Curto}, {Cuttaia}, {Davis}, {de Bernardis}, {de Rosa}, {de Zotti},
  {Delabrouille}, {Di Valentino}, {Dickinson}, {Diego}, {Dor{\'e}}, {Douspis},
  {Ducout}, {Dupac}, {Elsner}, {En{\ss}lin}, {Eriksen}, {Falgarone}, {Fantaye},
  {Finelli}, {Forastieri}, {Frailis}, {Fraisse}, {Franceschi}, {Frolov},
  {Galeotta}, {Galli}, {Ganga}, {G{\'e}nova-Santos}, {Gerbino}, {Ghosh},
  {Gonz{\'a}lez-Nuevo}, {G{\'o}rski}, {Gruppuso}, {Gudmundsson}, {Hansen},
  {Helou}, {Henrot-Versill{\'e}}, {Herranz}, {Hivon}, {Huang}, {Ili{\'c}},
  {Jaffe}, {Jones}, {Keih{\"a}nen}, {Keskitalo}, {Kisner}, {Knox},
  {Krachmalnicoff}, {Kunz}, {Kurki-Suonio}, {Lagache}, {L{\"a}hteenm{\"a}ki},
  {Lamarre}, {Langer}, {Lasenby}, {Lattanzi}, {Lawrence}, {Le Jeune},
  {Levrier}, {Lewis}, {Liguori}, {Lilje}, {L{\'o}pez-Caniego}, {Ma},
  {Mac{\'{\i}}as-P{\'e}rez}, {Maggio}, {Mangilli}, {Maris}, {Martin},
  {Mart{\'{\i}}nez-Gonz{\'a}lez}, {Matarrese}, {Mauri}, {McEwen}, {Meinhold},
  {Melchiorri}, {Mennella}, {Migliaccio}, {Miville-Desch{\^e}nes}, {Molinari},
  {Moneti}, {Montier}, {Morgante}, {Moss}, {Naselsky}, {Natoli}, {Oxborrow},
  {Pagano}, {Paoletti}, {Partridge}, {Patanchon}, {Patrizii}, {Perdereau},
  {Perotto}, {Pettorino}, {Piacentini}, {Plaszczynski}, {Polastri}, {Polenta},
  {Puget}, {Rachen}, {Racine}, {Reinecke}, {Remazeilles}, {Renzi}, {Rocha},
  {Rossetti}, {Roudier}, {Rubi{\~n}o-Mart{\'{\i}}n}, {Ruiz-Granados},
  {Salvati}, {Sandri}, {Savelainen}, {Scott}, {Sirri}, {Sunyaev}, {Suur-Uski},
  {Tauber}, {Tenti}, {Toffolatti}, {Tomasi}, {Tristram}, {Trombetti},
  {Valiviita}, {Van Tent}, {Vielva}, {Villa}, {Vittorio}, {Wandelt}, {Wehus},
  {White}, {Zacchei}, \& {Zonca}}]{2016A&A...596A.108P}
{Planck Collaboration}, {Adam}, R., {Aghanim}, N., {et~al.} 2016, \aap, 596,
  A108, \dodoi{10.1051/0004-6361/201628897}

\bibitem[{{Ryu} {et~al.}(2016){Ryu}, {Tanaka}, {Perna}, \&
  {Haiman}}]{2016MNRAS.460.4122R}
{Ryu}, T., {Tanaka}, T.~L., {Perna}, R., \& {Haiman}, Z. 2016, \mnras, 460,
  4122, \dodoi{10.1093/mnras/stw1241}

\bibitem[{{Shang} {et~al.}(2010){Shang}, {Bryan}, \&
  {Haiman}}]{2010MNRAS.402.1249S}
{Shang}, C., {Bryan}, G.~L., \& {Haiman}, Z. 2010, \mnras, 402, 1249,
  \dodoi{10.1111/j.1365-2966.2009.15960.x}

\bibitem[{{Smith} {et~al.}(2017){Smith}, {Becerra}, {Bromm}, \&
  {Hernquist}}]{2017MNRAS.472..205S}
{Smith}, A., {Becerra}, F., {Bromm}, V., \& {Hernquist}, L. 2017, \mnras, 472,
  205, \dodoi{10.1093/mnras/stx1993}

\bibitem[{{Smith} {et~al.}(2018){Smith}, {Regan}, {Downes}, {Norman}, {O'Shea},
  \& {Wise}}]{2018arXiv180406477S}
{Smith}, B., {Regan}, J., {Downes}, T., {et~al.} 2018, ArXiv e-prints.
\newblock \doarXiv{1804.06477}

\bibitem[{{Springel} {et~al.}(2001){Springel}, {Yoshida}, \&
  {White}}]{2001NewA....6...79S}
{Springel}, V., {Yoshida}, N., \& {White}, S.~D.~M. 2001, \na, 6, 79,
  \dodoi{10.1016/S1384-1076(01)00042-2}

\bibitem[{{Tanaka} \& {Haiman}(2009)}]{2009ApJ...696.1798T}
{Tanaka}, T., \& {Haiman}, Z. 2009, \apj, 696, 1798,
  \dodoi{10.1088/0004-637X/696/2/1798}

\bibitem[{{Tanaka} \& {Li}(2014)}]{2014MNRAS.439.1092T}
{Tanaka}, T.~L., \& {Li}, M. 2014, \mnras, 439, 1092,
  \dodoi{10.1093/mnras/stu042}

\bibitem[{{Taylor} \& {Kobayashi}(2014)}]{2014MNRAS.442.2751T}
{Taylor}, P., \& {Kobayashi}, C. 2014, \mnras, 442, 2751,
  \dodoi{10.1093/mnras/stu983}

\bibitem[{{Tseliakhovich} \& {Hirata}(2010)}]{2010PhRvD..82h3520T}
{Tseliakhovich}, D., \& {Hirata}, C. 2010, \prd, 82, 083520,
  \dodoi{10.1103/PhysRevD.82.083520}

\bibitem[{{Visbal} {et~al.}(2014){Visbal}, {Haiman}, \&
  {Bryan}}]{2014MNRAS.445.1056V}
{Visbal}, E., {Haiman}, Z., \& {Bryan}, G.~L. 2014, \mnras, 445, 1056,
  \dodoi{10.1093/mnras/stu1794}

\bibitem[{{Volonteri}(2010)}]{2010A&ARv..18..279V}
{Volonteri}, M. 2010, \aapr, 18, 279, \dodoi{10.1007/s00159-010-0029-x}

\bibitem[{{Volonteri} \& {Bellovary}(2012)}]{2012RPPh...75l4901V}
{Volonteri}, M., \& {Bellovary}, J. 2012, Reports on Progress in Physics, 75,
  124901, \dodoi{10.1088/0034-4885/75/12/124901}

\bibitem[{{Volonteri} {et~al.}(2003){Volonteri}, {Haardt}, \&
  {Madau}}]{2003ApJ...582..559V}
{Volonteri}, M., {Haardt}, F., \& {Madau}, P. 2003, \apj, 582, 559,
  \dodoi{10.1086/344675}

\bibitem[{{Volonteri} {et~al.}(2008){Volonteri}, {Lodato}, \&
  {Natarajan}}]{2008MNRAS.383.1079V}
{Volonteri}, M., {Lodato}, G., \& {Natarajan}, P. 2008, \mnras, 383, 1079,
  \dodoi{10.1111/j.1365-2966.2007.12589.x}

\bibitem[{{Wu} {et~al.}(2015){Wu}, {Wang}, {Fan}, {Yi}, {Zuo}, {Bian}, {Jiang},
  {McGreer}, {Wang}, {Yang}, {Yang}, {Thompson}, \&
  {Beletsky}}]{2015Natur.518..512W}
{Wu}, X.-B., {Wang}, F., {Fan}, X., {et~al.} 2015, \nat, 518, 512,
  \dodoi{10.1038/nature14241}

\end{thebibliography}

\end{document}